\documentclass[prb,twocolumn,showpacs,amsmath,superscriptaddress,psfig,amssymb]{revtex4-1}
\usepackage{amsmath}% Physical Review B
\usepackage{amssymb}
\usepackage{graphicx}
\usepackage{dcolumn}
\usepackage{bm}
\usepackage{txfonts}
\usepackage{cases}
\usepackage{bbding}

\begin{document}

\title{BCS-like superconductivity in noncentrosymmetric compounds Nb$_{x}$Re$_{1-x}$ (0.13$\leq x \leq$ 0.38)}
\author{J. Chen}
\affiliation{Center for Correlated Matter and Department of Physics ,
Zhejiang University, Hangzhou, Zhejiang 310027, China}
\author{L. Jiao}
\affiliation{Center for Correlated Matter and Department of Physics ,
Zhejiang University, Hangzhou, Zhejiang 310027, China}
\author{J. L. Zhang}
\affiliation{Center for Correlated Matter and Department of Physics ,
Zhejiang University, Hangzhou, Zhejiang 310027, China}
\author{Y. Chen}
\affiliation{Center for Correlated Matter and Department of Physics ,
Zhejiang University, Hangzhou, Zhejiang 310027, China}
\author{L. Yang}
\affiliation{Center for Correlated Matter and Department of Physics ,
Zhejiang University, Hangzhou, Zhejiang 310027, China}
\author{M. Nicklas}
\affiliation{Max Planck Institute for Chemical Physics of Solids,
D-01187 Dresden, Germany}
\author{F. Steglich}
\affiliation{Max Planck Institute for Chemical Physics of Solids,
D-01187 Dresden, Germany}
\author{H. Q. Yuan}
\email{hqyuan@zju.edu.cn} \affiliation{Center for Correlated Matter and Department of Physics ,
Zhejiang University, Hangzhou, Zhejiang 310027, China}
\date{\today}

\begin{abstract}

We present research on the superconducting properties of
Nb$_{x}$Re$_{1-x}$ ($x$ = 0.13-0.38) obtained by measuring the
electrical resistivity $\rho(T)$, magnetic susceptibility $\chi(T)$,
specific heat $C_P(T)$, and London penetration depth
$\Delta\lambda(T)$. It is found that the superconducting transition
temperature $T_c$ decreases monotonically with an increase of
$x$. The upper critical field $B_{c2}(T)$ for various $x$ can be
nicely scaled by its corresponding $T_c$. The electronic specific
heat $C_e(T)/T$, penetration depth $\Delta\lambda(T)$, and superfluid
density $\rho_{s}(T)$ demonstrate exponential behavior at low
temperatures and can be well fitted by a one-gap BCS model. The
residual Sommerfeld coefficient $\gamma_0(B)$ in the superconducting
state follows a linear field dependence. All these properties
suggest an \emph{s}-wave BCS-type of superconductivity with a very large $B_{c2}(0)$ for
Nb$_{x}$Re$_{1-x}$ (0.13 $\leq x \leq$ 0.38).
\end{abstract}


\maketitle

%\preprint{APS/123-QED}

% It is always \today, today,
%  but any date may be explicitly specified

% PACS, the Physics and Astronomy
% Classification Scheme.

\section{Introduction}

The discovery of superconductivity (SC) in the heavy fermion
compound CePt$_3$Si has attracted considerable interest in studying
the effects of broken inversion symmetry on superconducting pairing
states. \cite{Bauer04} While the inversion symmetry is absent in a
crystal, the resulting antisymmetric potential gradient causes a
parity-breaking antisymmetric spin-orbit coupling (ASOC), which may
lift up the spin degeneracy. An admixture of spin-singlet and
spin-triplet components is then allowed in the pairing states, whose
ratio might be tuned by the ASOC strength.
\cite{Gorkov01,Yip02,Frigeri04,Yanase07,Fujimoto07} Such a scenario
seems to be supported by the experimental observations of
Li$_2$(Pd$_{1-x}$Pt)$_3$B, in which evidence of BCS-like SC is shown
in Li$_2$Pd$_3$B, but a spin triplet was recognized in Li$_2$Pt$_3$B
with increasing ASOC strength. \cite{Yuan,Nishiyama07,Takeya07}
Furthermore, noncentrosymmetric (NCS) superconductors were recently
proposed as important candidates for studying topological SC.
\cite{Qi09,Lu10}

In the past few years, a growing number of NCS superconductors have
been studied, varying from heavy fermion compounds to a number
of weakly correlated intermetallic compounds. \cite{Book} Exotic
properties, including nodal SC, \cite{Yuan, Sigrist07, Mukuda10, Chen11}
multigap SC, \cite{Harada, Kuroiwa08, Klimczuk07, Chen12} and a huge
upper critical field, \cite{Kimura07, Settai08, Chen11} have been
experimentally observed in these systems. However, the role of
lacking inversion symmetry on the superconducting pairing states
remains a fundamentally open question. \cite{Chen12} Therefore, it
is highly desirable to elucidate the determining parameters of the
pairing state in NCS superconductors. Systematic studies on the
superconducting properties of NCS superconductors with a tunable ASOC
strength may help address these issues.

The intermetallic binary compound Nb$_{x}$Re$_{1-x}$ (0.13 $\leq x \leq$ 0.38) crystallizes in the cubic Ti$_5$Re$_{24}$-type structure
with a space group $I\overline{4}$3\emph{m} (No. 217), which loses
inversion symmetry on the sites of Nb(24\emph{g}) and
Re(24\emph{g}). SC was initially reported in these compounds by
Knapton \emph{et al}. in the 1950's. \cite{Knapton} Very recently, the
superconducting properties of Nb$_{0.18}$Re$_{0.82}$ were revisited as
an example of NCS superconductors; both specific heat and NMR
experiments indicated an $s$-wave type SC. \cite{Karki, Lue11} In order
to provide further characterizations of the superconducting pairing
state and its evolution as a function of $x$ in Nb$_{x}$Re$_{1-x}$,
we synthesized a series of polycrystals with varying Nb content, which are
expected to modify the ASOC strength due to the very different
atomic numbers of Nb and Re. The superconducting properties are
systematically studied by measuring the electrical resistivity
$\rho(T)$, magnetic susceptibility $\chi(T)$, specific heat
$C_p(T,B)$, as well as the London penetration depth
$\Delta\lambda(T)$. Our results provide strong evidence of BCS-like
SC for Nb$_{x}$Re$_{1-x}$.

\section{Experimental methods}

Polycrystalline Nb$_{x}$Re$_{1-x}$ ($x$ = 0.13-0.93) samples were
prepared by a two-step arc melting method in ultrapure argon gas. A
Ti button was used as an oxygen getter. In the first step, buttons
of high-purity niobium (99.99\%, Alfa Aesar) and rhenium (99.99\%,
Alfa Aesar) powders were prepared by arc melting, respectively. In
the second step, stoichiometric amounts of the two compositions were
melted together to form the alloy Nb$_{x}$Re$_{1-x}$. The ingot was
inverted and remelted several times to improve sample homogeneity.
Such a two-step approach is efficient to decrease the melting points
of materials through alloying. The so-derived ingot forms a hard
button with a negligible weight loss of less than 0.5\%, attributed
to the low vapor pressure of Nb and Re. A subsequent heat treatment
was performed at 800$^\circ$C in a vacuum-sealed quartz tube for 7
days, followed by a slow cooling of the furnace.

The crystal structure of the ingots was characterized by powder
x-ray diffractometry (XRD) using a X'Pert PRO diffractometer (Cu
$K\alpha$ radiation) in the Bragg-Brentano geometry for the
2$\Theta$ range of 10$^\circ$-90$^\circ$. Sample compositions were
identified by using energy dispersive x-ray (EDX) spectroscopy,
showing nearly the same compositions as the nominal values. The
electrical resistivity was measured using a standard four-probe
method in a \emph{dc} magnetic field up to 14T and at temperatures
down to 0.3K in a $^3$He cryostat. Measurements of the magnetic
susceptibility and specific heat were performed in a commercial
magnetic property measurement system (5T-MPMS) and a physical
property measurement system (9T-PPMS) (Quantum Design),
respectively. Measurements of the London penetration depth
$\Delta\lambda(T)$ were performed by using a technique based on a
tunnel diode oscillator (TDO) \cite{TDO} at a frequency of 7 MHz down
to 0.4 K in a $^3$He cryostat.

\section{Results and discussion}
\subsection{Crystal structures}

\begin{figure}[b]
\centering
\includegraphics[width=8.0cm]{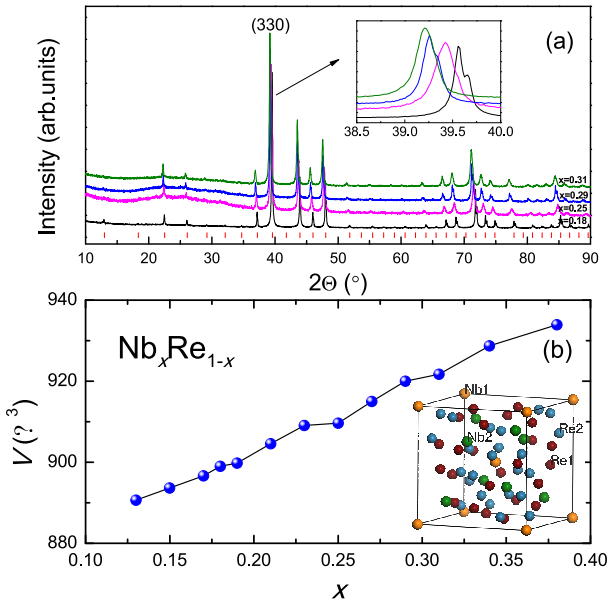}
\caption{(Color online) (a) XRD patterns for Nb$_{x}$Re$_{1-x}$,
$x$ = 0.18, 0.25, 0.29, and 0.38. Short vertical bars are the standard
reflection positions. (b) The unit cell volume plotted as a function of
$x$ for samples with $x$ = 0.13-0.38. The inset shows the
Ti$_5$Re$_{24}$-type crystal structure.} \label{fig1}
\end{figure}

Powder XRD spectra were taken on the Nb$_{x}$Re$_{1-x}$ samples with
various $x$ ($x$ = 0.13-0.93) at room temperature. It was found that
Nb$_{x}$Re$_{1-x}$ crystallizes in three different crystal structures
while varying the Nb composition from $x$ = 0.13 to $x$ = 0.93, which is
consistent with previous reports. \cite{Knapton} For 0.13 $\leq x \leq $ 0.38 (region I), XRD patterns
identify our samples as being of single phase with a cubic
Ti$_5$Re$_{24}$ structure [see Fig. 1(a)]. In each unit cell,
there are 58 atoms on four crystallographically
distinct sites: Nb(1), Nb(2), Re(1), and Re(2) [see the inset of Fig.
1(b)]. The inversion symmetry is preserved for the Nb(1) sites
(2\emph{a}), occupied by two Nb atoms, while it is broken on the
Nb(2), Re(1), and Re(2) sites along all crystallographic directions,
which are occupied by 8 Nb, 24 Re, and 24 Re atoms, respectively. For
0.38 $\leq x \leq$ 0.52 (region II), the Nb$_7$Re$_8$-type phase dominates
(tetragonal, $P$4$_2$/\emph{mnm}). However, XRD patterns reveal a
Nb structure with cubic $I$\emph{m}$\overline{3}$\emph{m} space
group (No. 229) in the range of 0.55 $\leq$ $x \leq$ 0.93 (region III),
demonstrating a maximum solubility of 46 \% rhenium in niobium.
\cite{Knapton} In regions I and II, there is a systematic shift of
the diffraction peaks to lower angles with increasing $x$,
suggesting a binary solid solution for Nb$_{x}$Re$_{1-x}$ ($x$ =
0.13-0.54). It is noted that the Nb$_{x}$Re$_{1-x}$ alloys are mixed with Nb$_5$Re$_{24}$
and Re-element phases for 0 $< x <$ 0.13. \cite{Knapton}
In this paper, we will mainly focus on the properties of the NCS superconductors in region I.

In Fig. 1(b), we plot the unit cell volume $V$ as a function of $x$
for Nb$_{x}$Re$_{1-x}$ (0.13 $\leq x \leq$ 0.38), as determined from
the XRD patterns using Rietveld refinement. One can see that the
unit cell volume increases monotonically with increasing $x$,
attributed to the larger atomic radius of Nb. We find a lattice
parameter of $a$ = 9.6507$\AA$ for Nb$_{0.18}$Re$_{0.82}$ which is
consistent with that reported in Ref. [23].

\subsection{The dependence of $T_c$ on Nb concentration $x$ }

\begin{figure}[b]
\centering
\includegraphics[width=8.0cm]{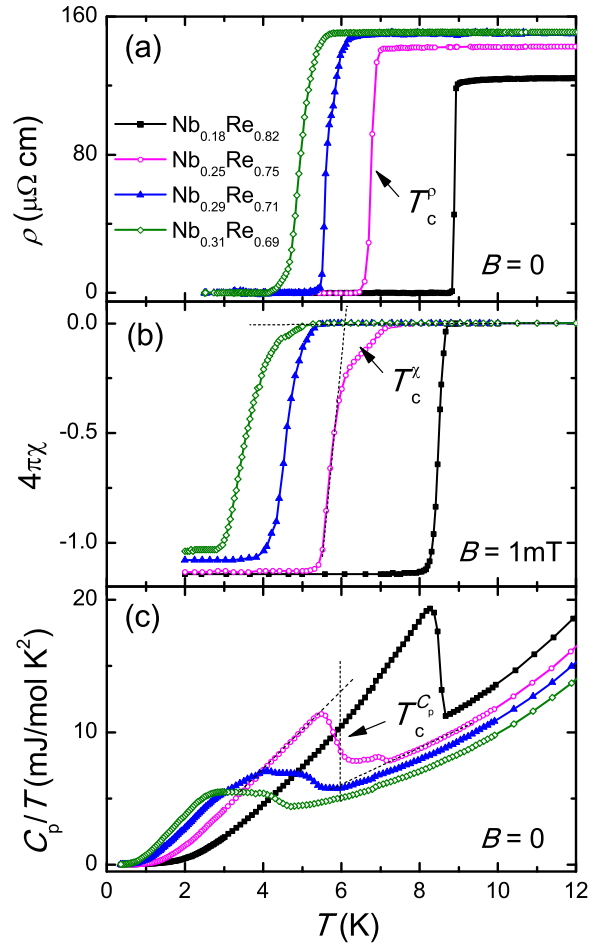}
\caption{(Color online) Temperature dependence of (a) the electrical
resistivity $\rho(T)$, (b) the magnetic susceptibility $\chi(T)$ (b), and
(c) the specific heat $C_p(T)$ for Nb$_{x}$Re$_{1-x}$ ($x$ = 0.18, 0.25,
0.29, and 0.31). Dashed lines illustrate the methods of determining
$T_c$.} \label{fig2}
\end{figure}

Figure 2 presents the temperature dependence of the electrical
resistivity $\rho(T)$ (a) [Fig. 2(a)], magnetic susceptibility $\chi(T)$ (b) [Fig. 2(b)], and
specific heat $C_p(T)$ (c) [Fig. 2(c)] for Nb$_{x}$Re$_{1-x}$ ($x$ = 0.18, 0.25,
0.29, and 0.31). A pronounced superconducting transition is observed
in all these quantities for each compound. Note that the
superconducting transition is broadened upon increasing $x$, which
is likely attributed to the enhanced sample inhomogeneity. Bulk SC with nearly 100\% shielding volume can be
inferred from the magnetic susceptibility as well as the specific-heat jumps at $T_c$. The Sommerfeld coefficient $\gamma_n$ can be
derived from the polynomial fits of the normal state specific heat
by $C_p$ = $\gamma_nT$ + ($B_3T^3$ + $B_5T^5$ + $B_7T^7$), in which the
first term represents the electronic contribution $C_e$ =
$\gamma_nT$, while the second term denotes the phonon contribution.
With this method, we obtain $\gamma_n$ = 4.8mJ/mol\textperiodcentered
K$^2$ for $x$ = 0.18, the $\gamma_n$ value decreasing with increasing
$x$. It is pointed out that a much larger value of $\gamma_n$ =
53.5mJ/mol\textperiodcentered K$^2$ was previously reported in Ref.
[23] for Nb$_{0.18}$Re$_{0.82}$, presumably due to a miscalculation.
Furthermore, no evidence of magnetic order and magnetic impurities
is observed in the above measurements.

In Fig. 3, we plot the superconducting transition temperatures
$T_c$ and the Sommerfeld coefficient $\gamma_n$ as a function of
the Nb content $x$ for Nb$_{x}$Re$_{1-x}$. Here we determine $T_c$ from
the intersections of the magnetic susceptibility, and the mid points
of the resistive transition and the specific heat jumps as
illustrated in Fig. 2. One can see that the bulk $T_c$s, derived
from the magnetic susceptibility and specific heat, are slightly
reduced but still compatible with the resistive $T_c$, showing a
monotonic decrease with increasing $x$. Such dependence of $T_c(x)$
may originate from the decrease of the density of states at the
Fermi level, as reflected by the $x$-dependence of $\gamma_n(x)$
in Fig. 3(b).

\begin{figure}[b]
\centering
\includegraphics[width=8.0cm]{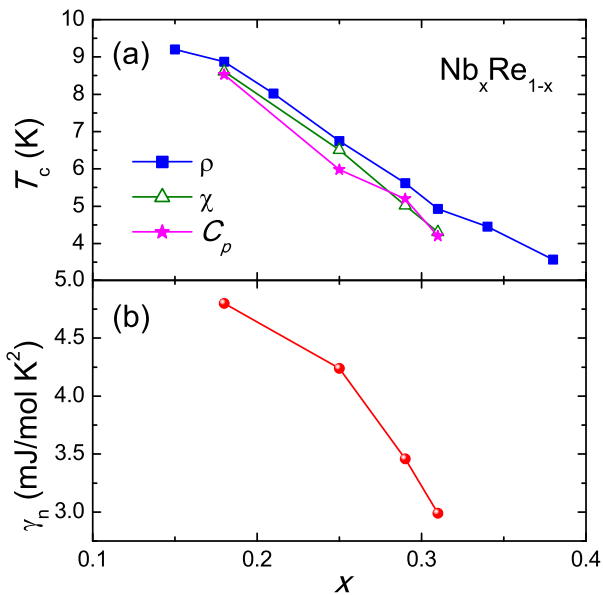}
\caption{(Color online) (a) $T_c$ vs $x$ for Nb$_{x}$Re$_{1-x}$, determined from the electrical resistivity $\rho(T)$,
the magnetic susceptibility $\chi(T)$, and the specific heat $C_p(T)$,
respectively. (b) The Sommerfeld coefficient $\gamma_n(x)$ plotted
as a function of $x$.} \label{fig3}
\end{figure}

\subsection{Upper critical field}

To determine the upper critical field $B_{c2}(T)$, we have measured
the electrical resistivity $\rho(T)$ ($x$ = 0.18, 0.25, 0.29, and 0.31)
and specific heat $C_p(T)$ ($x = 0.18$) at various magnetic fields. As
an example, we show $\rho(T)$ of Nb$_{0.18}$Re$_{0.82}$ at magnetic
fields up to 14 T in the inset of Fig. 4. Obviously, the
superconducting transition is shifted to lower temperatures upon
increasing magnetic field, but not yet suppressed by the maximum
field of 14 T we applied. The superconducting transition of $x = 0.18$
remains fairly sharp in a magnetic field, but it is broadened in
other samples with larger $x$, indicating that the samples become
more inhomogeneous with increasing Nb content. Similar behavior is
also seen in the specific-heat data (see below).

In the main panel of Fig. 4, we show the normalized upper critical
field $B_{c2}/[T_c(dB_{c2}/dT)_{T_c}]$ versus $T/T_c$ for various
Nb contents. Here the resistive $T_c$ and the horizontal bars are
determined from the mid-points, as well as the 10\% and 90\% drops
of the normal-state resistivity just above $T_c$. The values of
$T_c$ from the specific heat are determined by using the entropy
balance method in a plot of $C/T$ vs $T$. Remarkably, the upper critical fields $B_{c2}(T)$,
derived either from the same measurement of different sample
concentrations $x$ or from different measurements of the same sample
($x = 0.18$), can be nicely scaled by $T_c$; the normalized curves
collapse onto a single line as shown in Fig. 4.

\begin{figure}[b]
\centering
\includegraphics[width=8.0cm]{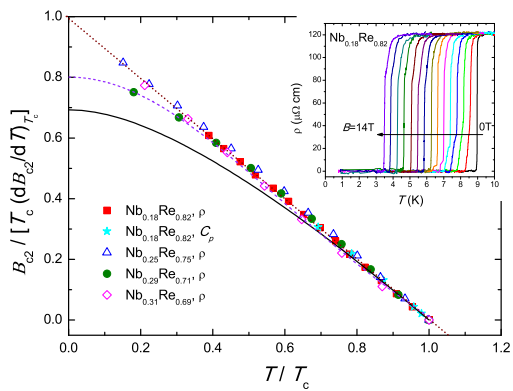}
\caption{(Color online) Normalized upper critical field
$B_{c2}/[T_c(dB_{c2}/dT)_{T_c}]$ vs $T/T_c$ for
Nb$_{x}$Re$_{1-x}$ ($x$ = 0.18, 0.25, 0.29, and 0.31). The solid and
dashed lines represent fits of the WHH and GL methods, respectively.
The inset shows the electrical resistivity $\rho(T)$ of
Nb$_{0.18}$Re$_{0.82}$ at various magnetic fields.} \label{fig4}
\end{figure}

Another important feature of the upper critical field curve $B_{c2}(T)$
is the remarkably linear temperature dependence down to the base
temperature $T$ $\simeq$ 0.3K. For comparison, in Fig. 4 we include
the fits of the upper critical fields by the Werthamer- Helfand-
Hohenberg (WHH) method in the dirty limit (solid line) \cite{WHH} as
well as the Ginzburg-Landau (GL) formula, \cite{GL}
$B_{c2}(T)=B_{c2}(0)[1-(T/T_{c})^{2}]/[1+(T/T_{c})^{2}]$ (dashed
line). One can see that the WHH method fails to describe the
experimental data over a wide temperature region. The GL formula can
give a much better illustration of the experimental data, but also
shows deviations at low temperatures. We estimate the upper critical
field $B_{c2}(0)$ by linearly extrapolating $B_{c2}(T)$ to zero
temperature, which gives $B_{c2}(0)$ = 23T, 14.5T, 11.8T, and 9T for
$x$ = 0.18, 0.25, 0.29, and 0.31, respectively. Such values of
$B_{c2}(0)$ largely exceed the corresponding orbital limiting
field, and are close to or even larger than the Pauli limiting field.

According to the WHH theory, \cite{WHH} the upper critical field
limited by the orbital mechanism can be estimated from $T_c$ and the
initial slope of the upper critical field $B_{c2}(T)$, i.e.,
\begin{equation}
B_{c2}^{orb}(0)= -0.69T_c(dB_{c2}/dT)_{T=T_c}.
\end{equation}
The above formula gives $B_{c2}^{orb}(0)$ = 16.1T, 10.5T, 8.6T, and
6.3T for $x$ = 0.18, 0.25, 0.29, and 0.31, respectively.

On the other hand, SC can be destroyed by the Pauli paramagnetic
effect in a magnetic field as a result of the Zeeman effect. The Pauli
limiting field is usually defined by: \cite{Pauli1,Pauli2}
\begin{equation}
 B_{c2}^{P}(0)= \Delta_0/\sqrt{2}\mu_B,
 \end{equation}
where $\Delta_0$ is the energy gap amplitude at zero temperature.
For a conventional BCS superconductor, $\Delta_0$ = 1.76$T_c$, Eq. 2
can be simplified as: $B_{c2}^{P}(0)$ = 1.86$T_c$. In the following
section, we will show that Nb$_x$Re$_{1-x}$ is a type of
weak-coupling BCS superconductor. The Pauli limiting fields are,
therefore, estimated to be 16.6T, 12.6T, 10.4T, and 9.1T for $x$ =
0.18. 0.25, 0.29, and 0.31, respectively.

The linear temperature dependence of $B_{c2}(T)$ and the absence of
Pauli limiting behavior are unusual for a BCS-type superconductor.
However, similar behavior was also observed in other
weakly correlated NCS superconductors. For example, a very large
upper critical field $B_{c2}(0)$ was also reported in the multigap
superconductors La$_2$C$_3$ (Ref. 30) and Y$_2$C$_3$.
\cite{Chen11} In the latter case, evidence of line nodes was
observed in the low-temperature limit. \cite{Chen11}
In the following, we will provide further experimental facts to reveal the
gap symmetry in Nb$_{x}$Re$_{1-x}$ by measuring the low-temperature
specific heat at various magnetic fields $C_p(T,B)$ as well as the
London penetration depth $\Delta \lambda(T)$. Since the sample
Nb$_{0.18}$Re$_{0.82}$ shows the highest sample quality among this
series of compounds, we will focus on this concentration here,
although other samples demonstrate similar behavior.

\subsection{Gap symmetry}

\subsubsection {Specific heat}

\begin{figure}[b]
\centering
\includegraphics[width=8.0cm]{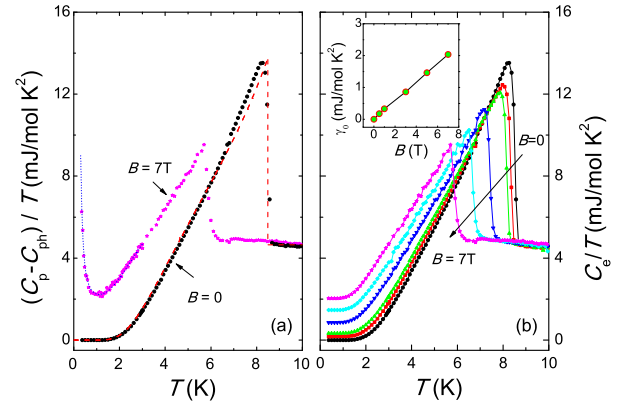}
\caption{(Color online) Temperature dependence of the specific heat at various magnetic fields
for Nb$_{0.18}$Re$_{0.82}$. (a) The
specific heat $C_p$ obtained after subtracting the phonon contributions from the raw data at
$B$ = 0 and 7 T. The dashed line shows a fit to the BCS model with $\Delta_0$ = 1.93$T_c$. The dotted line denotes a
power-law fit of the nuclear Schottky contributions $C_{\mathrm{Sch}}(T,B)$
at low temperatures. (b) The electronic specific heat $C_e(T)$ at
$B$ = 0, 0.5, 1, 3, 5, and 7 T. The inset plots the magnetic field
dependence of the residual Sommerfeld coefficient $\gamma_0(B)$.}
\label{fig5}
\end{figure}

The temperature dependence of the specific heat $C_p(T)$ was
previously reported for Nb$_{0.18}$Re$_{0.82}$ at zero field by
Karki \emph{et al}. \cite {Karki} It shows exponential behavior in
the low-temperature limit, suggesting BCS-type SC with a moderate
electron-phonon coupling. In our measurements, similar behavior is
observed for Nb$_x$Re$_{1-x}$ [see Figs. 2(c) and 5(a)]. Upon
increasing the Nb content from $x$ = 0.18 to 0.31, the specific-heat
jump $\Delta C/\gamma_nT_c$ at $T_c$ varies from 1.86 to about 0.5.
Note that the broadened superconducting transitions for samples with
large $x$ do not allow us to reliably determine the specific heat
jumps accurately enough. Far below the superconducting transition
($T$ $<$ 0.3$T_c$), the electronic specific heat $C_e(T)/T$ for
various Nb contents shows a very weak temperature dependence which
can be reasonably fitted by the BCS-type exponential behavior.
Following the procedures described in Ref. [19], in Fig. 5(a) we fit
the specific heat $C_e(T)/T$ of Nb$_{0.18}$Re$_{0.82}$ from the base
temperature up to $T_c$, in which $C_e(T)/T$ data can be nicely
described by the one-gap BCS model with an energy gap of $\Delta_0$ =
1.93$T_c$.

We further characterize the superconducting pairing state of
Nb$_{0.18}$Re$_{0.82}$ by measuring the specific heat in a magnetic
field. A pronounced upturn in $C_p(T)/T$ appears at low temperatures,
which shifts to higher temperatures upon increasing magnetic field.
This is ascribed to the high-$T$ tail of a nuclear Schottky anomaly.
In this case, the total specific heat can be expressed as
$C_p(T,B)$ = $C_e(T,B)$ + $C_{\mathrm{ph}}(T)$ + $C_{\mathrm{Sch}}(T,B)$, where
$C_e(T,B)$, $C_{\mathrm{ph}}(T)$, and $C_{\mathrm{Sch}}(T,B)$ represent the electron,
phonon, and nuclear Schottky contributions, respectively. The phonon
contributions can be subtracted from the polynomial fits above $T_c$
as illustrated in Sec. III B. The nuclear Schottky contribution
follows the expression of $C_{\mathrm{Sch}}(T,B)$ = $aB^2/T^2$, which can be
subtracted by fitting the low-temperature specific-heat data [see
Fig. 5(a), data for $B$ = 7 T]. In Fig. 5(b), we plot the temperature
dependence of the electronic specific heat $C_e(T)$ at various
magnetic fields for Nb$_{0.18}$Re$_{0.82}$, as obtained after
subtraction of the phonon and nuclear contributions. One can see
that application of a magnetic field eventually shifts the sharp
superconducting transition to lower temperatures and enhances the
residual Sommerfeld coefficient $\gamma_0(B)$. Note that the
specific-heat data at zero field show an extremely small residual
Sommerfeld coefficient of $\gamma_0$ = 0.018mJ/mol K$^2$, confirming
the good quality of the sample. The inset of Fig. 5(b) presents the
field dependence of the residual Sommerfeld coefficient
$\gamma_0(B)$, which shows a remarkably linear field dependence,
providing strong evidence of BCS-type SC for Nb$_{0.18}$Re$_{0.82}$.
In fully gapped superconductors, the low-lying excitations are
usually confined to the vortex cores, and the specific heat is
proportional to the vortex density which increases linearly with
increasing magnetic field, i.e., $\gamma_0(B)$ $\sim$ $B$.
\cite{Caroli}

\subsubsection{London penetration depth}

\begin{figure}[b]
\centering
\includegraphics[width=8.0cm]{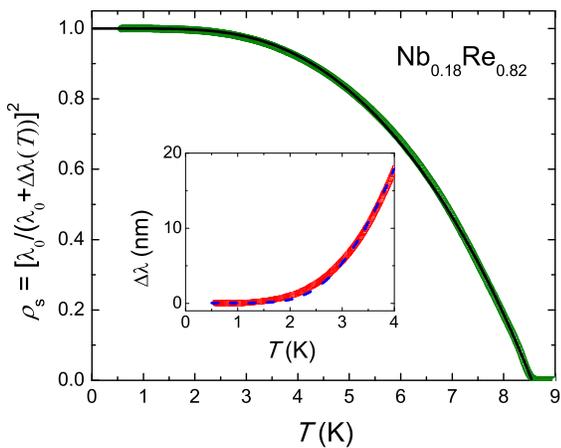}
\caption{(Color online) Temperature dependence of the superfluid
density $\rho_s(T)$ for Nb$_{0.18}$Re$_{0.82}$. The inset shows the penetration depth
$\Delta\lambda(T)$ at low temperatures. The lines represent fits to the BCS model.} \label{fig6}
\end{figure}

While the specific heat $C_p(T)/T$ explores the quasiparticle density
of states, the London penetration depth probes the superfluid density $\rho_s(T)$. In
this section, we present a precise measurement of the penetration
depth changes $\Delta\lambda(T)$ as well as $\rho_s(T)$ for
Nb$_{0.18}$Re$_{0.82}$ by using a TDO-based technique.
With this method, the change of the London penetration depth
$\Delta\lambda(T)$ is proportional to the resonant frequency shift
$\Delta f(T)$, i.e., $\Delta\lambda(T)$ = $G$ $\Delta f(T)$,
where the $G$ factor is a constant, solely determined by the
sample and coil geometries. \cite{TDO} The inset of Fig. 6
shows the penetration depth $\Delta\lambda(T)$ at low temperatures for
Nb$_{0.18}$Re$_{0.82}$, where $G$ = 1.4 nm$/$Hz. A
sharp superconducting transition with $T_c^{TDO}$ = 8.7 K is observed
in this sample (not shown), which is highly consistent with all the other
measurements. The penetration depth exhibits very weak temperature
dependence at low temperatures. According to the isotropic BCS
model, the penetration depth $\Delta\lambda(T)$ can be approximated
by the following exponential temperature dependence at $T$ $\ll$
$T_c$: $\Delta\lambda(T)=\lambda(T)-\lambda_0=
\lambda_0\sqrt{\frac{\pi\Delta_0}{2T}}e^{-\frac{\Delta_0}{T}}$,
where $\lambda_0$ and $\Delta_0$ are the penetration depth and gap
amplitude at zero temperature, respectively. In the inset of Fig. 6, the dashed line shows a fit of the BCS model to our experimental data $\Delta\lambda(T)$. Here we fix $\lambda_0$ = 414 nm as estimated below, and the derived energy gap of $\Delta_0$ = 1.91$T_c$ is remarkably consistent with that from the specific-heat data.

To further analyze the superconducting gap symmetry, in the following we turn to the superfluid density $\rho_s(T)$ of
Nb$_{0.18}$Re$_{0.82}$. The normalized superfluid density can be
converted from the penetration depth by $\rho_s(T)$ =
$[\lambda_0/\lambda(T)]^2$. Here we estimate the value of $\lambda_0$ according to the BCS and Ginzburg-Landau theories for a
type-II superconductor, \cite{Gross} i.e.,
$\lambda_0=\frac{1}{1.76T_c}\sqrt{\frac{\Phi_0B_{c2}(0)}{24\gamma_n}}$, where
$\Phi_0$ is the magnetic flux quantum. By taking the parameters derived from our specific-heat
data, i.e., $T_c^{C_p}$ = 8.5K, $B_{c2}^{C_p}(0)$ = 22.85T, and $\gamma_n$=
4.8 mJ/mol K$^2$ =
0.51$\times$10$^4$ erg/cm$^3$ K$^2$, we derived $\lambda_0$=
414 nm for Nb$_{0.18}$Re$_{0.82}$, which is compatible with that
obtained from the measurements of the lower critical field. \cite{Karki} In the main panel of Fig. 6, we plot the
temperature dependence of the superfluid density $\rho_s(T)$ for
Nb$_{0.18}$Re$_{0.82}$.

In order to adopt a proper model to fit the superfluid density $\rho_s(T)$ , we estimated the mean free path ($l$ $\approx$ 5 nm) and the coherence length ($\xi_0$ $\approx$ 4 nm) of the Nb$_{0.18}$Re$_{0.82}$ sample from the resistivity $\rho(T_c)$ = 120 $\mu\Omega$ cm, and the above mentioned quantities of $T_c^{C_p}$, $B_{c2}^{C_p}(0)$, and $\gamma_n$. \cite{Orlando} The close values of the mean free path and the coherence length suggest that it is appropriate to treat the sample in the dirty limit. Accordingly, we analyze our superfluid density $\rho_s(T)$ in terms of the \emph{s}-wave weak-coupling BCS model in the dirty limit, \cite{GL} i.e.,
\begin{equation} \rho_s(T)=\frac{\Delta(T)}{\Delta_0}\tanh(\frac{\Delta(T)}{2T}).
\label{eq:three}
\end{equation}
Here the temperature dependence of the gap function is given by \cite{Prozorov}
\begin{equation}
\Delta(T)=\Delta_0\tanh[\frac{\pi T_c}{\Delta_0}\sqrt{a\frac{\Delta
C}{C_e}(\frac{T_c}{T}-1)}], \label{eq:four}
\end{equation}
where $\Delta C/C_e$ is the relative jump in the electronic specific heat at
$T_c$ and $a$ = 2/3 for an isotropic BCS superconductor. As shown in Fig. 6, the experimental data $\rho_s(T)$ of Nb$_{0.18}$Re$_{0.82}$ can be well described by the BCS model with a gap magnitude of $\Delta_0$ = 1.95$T_c$, which agrees well with the
values derived from our fits of the specific-heat, and penetration depth data, as
well as the previous NMR data. \cite{Lue11} These measurements clearly
identify Nb$_{0.18}$Re$_{0.82}$ as an \emph{s}-wave BCS superconductor.
We note that similar results were also obtained for
Nb$_{0.29}$Re$_{0.71}$, suggesting BCS-like SC
for Nb$_{x}$Re$_{1-x}$ (0.13 $\leq$ $x \leq$ 0.38).

\section{Conclusion}

In summary, we have studied the superconducting properties of the
NCS compounds Nb$_x$Re$_{1-x}$ (0.13 $\leq$ $x \leq$ 0.38) by
measuring the electrical resistivity, magnetic susceptibility,
specific heat, as well as the London penetration depth. We found
that the superconducting transition temperature $T_c$ decreases
monotonically with increasing $x$, which is partially attributed to
the decrease of the density of states at the Fermi energy. The upper
critical field $B_{c2}(T)$ of Nb$_{x}$Re$_{1-x}$ can be perfectly
scaled by the corresponding $T_c$, showing a rather linear
temperature dependence down to the base temperature. The upper
critical field at zero temperature $B_{c2}(0)$ exceeds the orbital
limit, and also approaches the Pauli paramagnetic limit. On the other hand, the
temperature dependence of the electronic specific-heat coefficient $C_e(T)/T$,
penetration depth $\Delta\lambda(T)$ and superfluid density
$\rho_s(T)$ of Nb$_{0.18}$Re$_{0.82}$ can be consistently described
by the BCS model with an energy gap of $\Delta_0$ $\approx$ 1.9$T_c$.
Evidence of BCS SC for $x$ = 0.18 is further provided by the
observation of a linear field dependence of the residual
Sommerfeld coefficient $\gamma_0(B)$. Our results demonstrate that Nb$_{x}$Re$_{1-x}$ (0.13 $\leq$ $x \leq$ 0.38) is an
\emph{s}-wave BCS-type superconductor with negligible contributions
from the spin-triplet component, in spite of the heavy atomic mass of Re
residing on the NCS sites. The observation of $s$-wave SC
with an extremely large $B_{c2}$(0) in Nb$_{x}$Re$_{1-x}$, which is rare among the NCS superconductors, demands further theoretical and experimental investigations.

\section{Acknowledgments}

We appreciate valuable discussions with L. Z. Sun and M. B. Salamon.
This work was supported by the National Basic Research Program of
China (Grants No. 2011CBA00103, and No. 2009CB929104), the Natural Science
Foundation of China (Grant No. 10934005), Zhejiang Provincial Natural
Science Foundation of China, the Fundamental Research Funds for the
Central Universities, and the Max Planck Society under the auspices
of the Max Planck Partner Group of the MPI for Chemical Physics of
Solids, Dresden.

\end{document}